\documentclass[aps,prl,twocolumn,groupedaddress,floatfix]{revtex4}
\usepackage{graphicx}  
\usepackage{dcolumn}   
\usepackage{bm}        
\usepackage{amssymb}   
\graphicspath{ {figure/} }
\usepackage{subfig}
\usepackage{float}
\begin{document}


\title{Absolute and non-invasive determination of the electron bunch length in a Free Electron Laser using a Bunch Compressor Monitor}


\author{Gian Luca Orlandi}
\email[corresponding author: ]{gianluca.orlandi@psi.ch}
\affiliation{Paul Scherrer Institut, Forschungsstrasse 111, 5232 Villigen PSI, Switzerland}


\date{\today}

\begin{abstract}
In a linac driven Free Electron Laser (FEL), the shot-to-shot and non-invasive monitoring of the electron bunch length is normally ensured by Bunch Compressor Monitors (BCMs). The bunch-length dependent signal of a BCM results from the detection and integration - over a given frequency band - of the temporal coherent enhancement of the radiation spectral energy emitted by the electron beam while experiencing a longitudinal compression. In this letter, we present a method that permits to express the relative variation of the bunch length as a function of the relative statistical fluctuations of the BCM signal and charge. Furthermore, in the case of a BCM equipped with two detectors simultaneously operating in two distinct wavelength bands, the method permits an absolute determination of the bunch length. The presented method is beneficial to a FEL since it permits to tune the machine compression feedback with respect to the measured bunch length instead of the bunch-length dependent signal. In a CW-linac driven FEL, it can offer the precious opportunity to implement a fully non-invasive and absolute diagnostics of the bunch length.
\end{abstract}


\maketitle
The lasing performance of a linac driven Free Electron Laser (FEL) strongly relies on the stability over time of the low emittance and high density current features that the electron beam can achieve after the different acceleration and compression stages operated in the machine \cite{LCLS,SACLA,PAL,EuropXFEL,atomic1,Milne,atomic2,atomic3,chemich1,chemich2,biological1,biological2}. At every compression stage, the non-invasive and shot sequential monitoring of the electron bunch length in a FEL can be ensured by Bunch Compressor Monitors (BCMs) \cite{Wu,Loos,Frei2,Frei,Lockmann,Gerth}. The bunch-length dependent signal of a BCM results from the detection and integration over a given frequency band of the temporal coherent threshold of the radiation spectral energy emitted by the electron beam while crossing the last dipole of a magnetic chicane \cite{Williams,Ishi} or the hole of a diffraction radiation screen placed in a straight section just downstream \cite{Happek,Shibata,Shibata2,Cast,Veronese}. Thanks to the non-invasive and shot sequential feature, a BCM signal can be fruitfully exploited to feed back the RF working point of the accelerator and stabilize the bunch compression during FEL operations. Three BCMs are in operation in SwissFEL \cite{Frei,SFCDR}, two of them are integrated in the machine feedback. The present work will show how the bunch-length dependent signal of the BCM can be suitably processed for the shot-to-shot tracking of the relative variation of the bunch length. Furthermore, in the case of a BCM equipped with two detectors simultaneously collecting the radiation in two different wavelength bands, the presented method permits an absolute determination of the electron bunch length from the analysis of the relative statistical fluctuations of both the charge and BCM signals. In SwissFEL \cite{SFCDR,TS}, a 28 ns long 2-bunch macro-pulse with a charge of 10 or 200 pC and a longitudinal length of about 3 ps is accelerated up to about 6 GeV at a repetition rate of 100 Hz and compressed down to about a few fs in two magnetic chicanes (BC1 and BC2). After a further acceleration, a magnetic switch yard \cite{Paraliev} splits the 2 bunches off into the hard x-ray and soft x-ray brunches: ARAMIS \cite{Prat} and ATHOS \cite{Abela}, respectively. A further bunch compression down to the sub-fs scale can be operated by means of a magnetic chicane (ECOL) just upstream of the ARAMIS undulator line. Both BC1- and ECOL-BCM detect the Edge Synchrotron Radiation (Edge-SR) from the front edge of the 4th dipole of the magnetic chicane \cite{Frei}. Just downstream of BC2, a holed diffraction screen constitutes the spectral radiation source of the BCM. The BC1-BCM \cite{Frei} -  optimized for bunch lengths 220-290 fs (rms) - is equipped with two broadband Schottky diodes which, being sensitive up to more than 2 THz, are simultaneously illuminated by the SR light pulse that two high-pass spectral THz filters in cascade - with low-frequency cut-off of 0.3 and 0.6 THz, respectively - split off into two distinct optical paths. The BC2-BCM being designed for electron bunch length 3-25 fs (rms) is equipped with a Mercury Cadmium Telluride (MCT) detector with a sensitivity in the wavelength band 2-12 $\mu m$ \cite{Frei}. The ECOL-BCM - optimized for bunch length measurements in the range 0.7-3.0 fs (rms) \cite{Frei} - is equipped with a pyrodetector and an optical fiber spectrometer covering the spectral wavelength band 0.9-4.0 $\mu m$ and 0.9-2.5 $\mu m$, respectively. Both the pyrodetector and the spectrometer are illuminated by the same light pulse split apart in reflection and transmission by a calcium fluoride ($CaF_2$) beam splitter. Upon integration over the acceptance solid angle $\Delta\Omega$ of the BCM detector, the spectral distribution of the radiation energy emitted per unit of angular frequency $\omega=2\pi\nu$ by a bunch of N electrons at the temporal coherent threshold of the N-quadratic enhancement reads:
\begin{eqnarray}
\frac{dI^{Ne}(\omega)}{d\omega}\simeq N(N-1)^2F(\omega)\frac{dI^e(\omega)}{d\omega}\simeq N^2F(\omega)\frac{dI^e(\omega)}{d\omega},\label{equ1}
\end{eqnarray}
where $\frac{dI^e(\omega)}{d\omega}$ is the single particle energy spectrum and $F(\omega)$ is the longitudinal form factor (FF) of the electron beam \cite{Nodvick,Hirschmugl}. The FF being defined as the square module of the Fourier transform of the density distribution of the N electron coordinates along the longitudinal direction reads $F(\omega)=e^{-(\frac{\omega\sigma}{c})^2}$ for a Gaussian beam with a length $\sigma$.
In the following, we will assume that the single electron energy spectrum $\frac{dI^e(\omega)}{d\omega}$ is either slowly dependent on or even independent of the frequency in the considered frequency band. Such an assumption is reasonably valid in the case of two Edge-SR based BCMs of SwissFEL \cite{Bosch,Bosch2}. Moreover, the presented method, being based on the processing of the relative statistical fluctuation of the BCM with respect to a reference value, is mainly sensitive to the bunch length dependent component of eq.(\ref{equ1}) rather than to those parameters which in eq.(\ref{equ1}) are bunch-length invariant in the given frequency band such as, for instance, a low frequency diffractive cutoff of the single particle radiation energy spectrum or the frequency response function of the detector. Therefore, possible bunch-length invariant factors affecting the radiation energy spectrum of the electron bunch tends to be smoothed down. According to such hypothesis, let's normalize the formula in eq.(\ref{equ1}) with respect to $\frac{dI^e(\omega)}{d\omega}$. Afterwards, let's calculate in both sides of the resultant normalized expression of eq.(\ref{equ1}) the integral over the acceptance frequency band $\Delta\omega=(\omega_{max}-\omega_{min})$ of the detector and call $I$ the integral of the "normalized" radiation energy spectrum of the electron bunch which corresponds to a given form factor $F(\omega)$:
\begin{eqnarray}
I=\int_{\omega_{min}}^{\omega_{max}}d\omega\left(\frac{dI^{Ne}(\omega)}{d\omega}/\frac{dI^e(\omega)}{d\omega}\right).\label{equ4}
\end{eqnarray}
Upon applying the natural logarithm to both side of the so obtained normalized and integrated expression of eq.(\ref{equ1}), we obtain the following equation:
\begin{eqnarray}
\ln(I)=2\ln(N)+\ln(\int_{\omega_{min}}^{\omega_{max}}d\omega F(\omega)),\label{equ3}
\end{eqnarray}
By differentiating eq.(\ref{equ3}), it is possible to express the shot-to-shot relative variation of the BCM signal $\frac{\Delta I}{I}=\frac{I^*-I}{I}$ as a function of the corresponding relative fluctuation of the longitudinal form factor $\frac{\Delta F}{F}=\frac{F^*-F}{F}$ - i.e., of the electron bunch length $\frac{\Delta\sigma}{\sigma}=\frac{\sigma^*-\sigma}{\sigma}$ - and of the relative fluctuation of the number of electrons $\frac{\Delta N}{N}=\frac{N^*-N}{N}$ in the bunch:
\begin{eqnarray}
\frac{\Delta I}{I}=2\frac{\Delta N}{N}+\frac{\int_{\omega_{min}}^{\omega_{max}}d\omega[F^*(\omega)-F(\omega)]}{\int_{\omega_{min}}^{\omega_{max}}d\omega F(\omega)}.\label{equ5}
\end{eqnarray}
Thanks to a Taylor series expansion at the first order in $\frac{\Delta\sigma}{\sigma}$, the FF variation in the integrand of eq.(\ref{equ5}) can be explicitly expressed as \cite{Orlandi}
\begin{eqnarray}
&&F^*(\omega)-F(\omega)=e^{-(\frac{\omega\sigma}{c})^2(1+\frac{\Delta\sigma}{\sigma})^2}-e^{-(\frac{\omega\sigma}{c})^2}\simeq\nonumber\\
&&\simeq-2\frac{\Delta\sigma}{\sigma}\left(\frac{\omega\sigma}{c}\right)^2e^{-(\frac{\omega\sigma}{c})^2}=-2\frac{\Delta\sigma}{\sigma}\left(\frac{\omega\sigma}{c}\right)^2F(\omega)\label{equ6}
\end{eqnarray}
and the integral of the FF in eq.(\ref{equ5}) explicitly calculated
\begin{eqnarray}
&&\frac{\int_{\omega_{min}}^{\omega_{max}}d\omega[F^*(\omega)-F(\omega)]}{\int_{\omega_{min}}^{\omega_{max}}d\omega F(\omega)}
\simeq\nonumber\\
&&\simeq-2\frac{\Delta\sigma}{\sigma}\frac{\left[\frac{-2\omega\sigma e^{-(\frac{\omega\sigma}{c})^2}+\sqrt{\pi}c\, erf\left(\omega\sigma/c\right)}{4\sigma} \right]_{\omega_{min}}^{\omega_{max}}}{\left[\frac{\sqrt{\pi}c}{2\sigma}erf(\omega\sigma/c)\right]_{\omega_{min}}^{\omega_{max}}},\label{equ7}
\end{eqnarray}
where erf(x) is indicating the error function \cite{GR}.
In conclusion, from eqs.(\ref{equ5},\ref{equ6},\ref{equ7}) the relative fluctuation of the BCM signal $\frac{\Delta I}{I}$ with respect to a reference value - for instance, the mean value over a temporal sequence of acquisitions - can be expressed as a function of the corresponding relative variations of the number of electrons in the bunch $\frac{\Delta N}{N}$ and of the electron bunch length $\frac{\Delta\sigma}{\sigma}$:
\begin{eqnarray}
\frac{\Delta I}{I}=2\frac{\Delta N}{N}+\frac{\Delta\sigma}{\sigma}G(\sigma,\Delta\omega),\label{equ8}
\end{eqnarray}
where
\begin{eqnarray}
G(\sigma,\Delta\omega)=\left\{\frac{2\sigma}{\sqrt{\pi}c}\frac{\left[e^{-(\frac{\omega\sigma}{c})^2}\omega\right]_{\omega_{min}}^{\omega_{max}}}{\left[erf(\omega\sigma/c)\right]_{\omega_{min}}^{\omega_{max}}}-1\right\}.\label{equ9}
\end{eqnarray}
According to eqs.(\ref{equ8},\ref{equ9}), the simultaneous determination of $\sigma$ and $\frac{\Delta\sigma}{\sigma}$ seems to be not straightforward from the analysis of the signal of a BCM equipped with a single detector. Nevertheless, provided that the reference BCM signal has been calibrated in advance by means of an absolute monitor - for instance, a Transverse Deflecting Structure (TDS) \cite{Emma,TDC1,TDC2,TDC3,TDSInj} - eqs.(\ref{equ8},\ref{equ9}) permits a shot-to-shot tracking of the relative variation of the bunch length $\frac{\Delta\sigma}{\sigma}$ as a function of the relative fluctuations of the BCM and charge monitor signals: $\frac{\Delta I}{I}$ and $\frac{\Delta N}{N}$.
\begin{figure}[!tbh]
    \centering
    \includegraphics[width=0.4\textwidth]{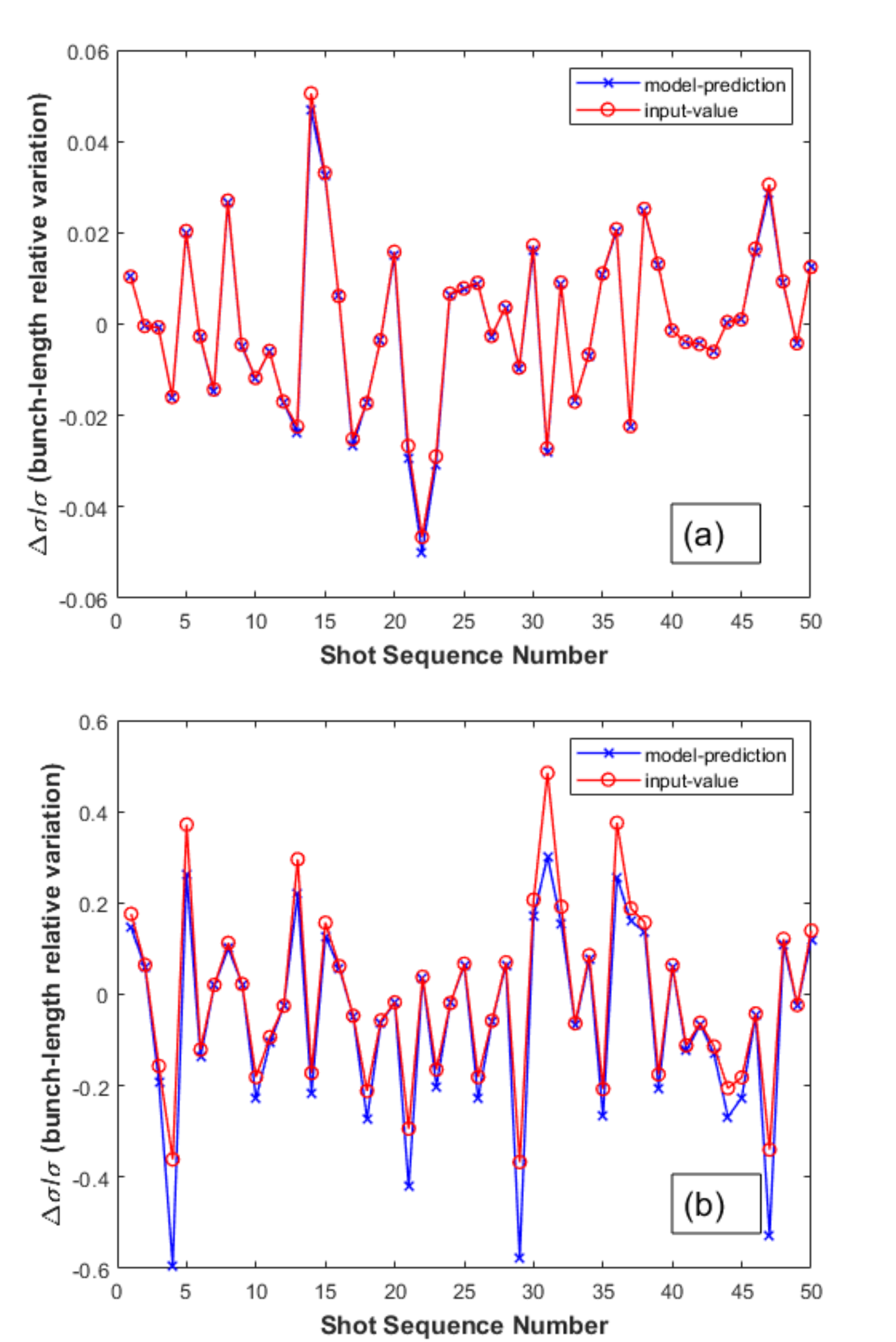}
    \caption{BC1-BCM ($0.3-2.0$ THz) and Gaussian FF with $\sigma=270$ fs: model-predicted $\frac{\Delta\sigma}{\sigma}$ vs model-input $\frac{\Delta\sigma}{\sigma}$ with rms deviations of 2$\%$ (a) and 20$\%$ (b); rms deviation of 1$\%$ for $\frac{\Delta N}{N}$.}
    \label{Orlandi:fig1}
\end{figure}
In the case of the BC1-BCM  with an acceptance frequency band of $0.3-2.0$ THz, the capability of the eqs.(\ref{equ8},\ref{equ9}) to predict $\frac{\Delta\sigma}{\sigma}$ was numerically tested for a Gaussian FF with $\sigma=270$ fs, see fig.(\ref{Orlandi:fig1}). A sequence of 500 relative variations of the bunch length $\frac{\Delta\sigma}{\sigma}$ around the reference value was randomly extracted according to a normally distributed generator with rms deviations from 0.5$\%$ up to 25$\%$. With the same random extraction procedure, a sequence of 500 relative variations of the beam charge $\frac{\Delta N}{N}$ was also obtained for rms deviations 0.5, 1 and 2$\%$.
From the randomly extracted sequences of $\frac{\Delta\sigma}{\sigma}$, the corresponding relative variation $\frac{\Delta I}{I}$ of the BCM signal was obtained via eqs.(\ref{equ1},\ref{equ4}). The so obtained sequences of $\frac{\Delta N}{N}$ and $\frac{\Delta I}{I}$ were used as input variables of the eqs.(\ref{equ8},\ref{equ9}) to evaluate the relative variation of the bunch length $\frac{\Delta\sigma}{\sigma}$. In fig.(\ref{Orlandi:fig1}), a sequence of model-predicted values of $\frac{\Delta\sigma}{\sigma}$ - eqs.(\ref{equ8},\ref{equ9}) - is compared with homologous model-input sequences randomly generated with rms deviations 2$\%$ and 20$\%$ for a charge fluctuation sequence $\frac{\Delta N}{N}$ with rms deviation of 1$\%$. In fig.(\ref{Orlandi:fig2}), the standard deviations of the 500-shot sequences of the model-predicted and model-input $\frac{\Delta\sigma}{\sigma}$ are plotted for different rms deviations of $\frac{\Delta N}{N}$. The model-predicted and model-input sequences of $\frac{\Delta\sigma}{\sigma}$ remain reasonably aligned up to a rms deviation of 10$\%$, see fig.(\ref{Orlandi:fig2}).
\begin{figure}[!tbh]
    \centering
    \includegraphics[width=0.4\textwidth]{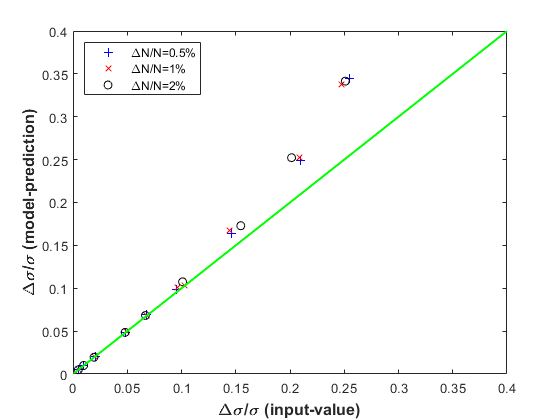}
    \caption{Standard deviations of model-predicted $\frac{\Delta\sigma}{\sigma}$ vs model-input reference values for $\frac{\Delta N}{N}$ with rms deviations 0.5-2$\%$. See also caption fig.(\ref{Orlandi:fig1}).}
    \label{Orlandi:fig2}
\end{figure}

For a BCM equipped with two independent detectors simultaneously integrating in two distinct frequency bands $\Delta\omega_{i}=(\omega_{max}^{i}-\omega_{min}^{i})$ - with $i=1,2$ - the same radiation pulse split off by a beam splitter, $\left(\frac{\Delta I}{I}\right)_i$, an absolute determination of the electron bunch length $\sigma$ is instead possible via eqs.(\ref{equ8},\ref{equ9}) by means of the following formula:
\begin{eqnarray}
\left[\left(\frac{\Delta I}{I}\right)_2-2\frac{\Delta N}{N}\right]=&&\frac{G(\sigma,(\Delta\omega)_2)}{[G(\sigma,(\Delta\omega)_2)-G(\sigma,(\Delta\omega)_1)]}\times\nonumber\\
&&\times\left[\left(\frac{\Delta I}{I}\right)_2-\left(\frac{\Delta I}{I}\right)_1\right].\label{equ10}
\end{eqnarray}
The complex functional dependence of eq.(\ref{equ10}) prevents a straightforward inversion of the formula with respect to $\sigma$. An easy way to handle eq.(\ref{equ10}) and determine $\sigma$ consists in: (a) applying a standard deviation operator to the eq.(\ref{equ10}) when running over the measured sequences of relative variations of BCM signals $\left(\frac{\Delta I}{I}\right)_i$ (with $i=1,2$) and charge monitor readout $\frac{\Delta N}{N}$; finally, (b) finding the "zeros" of the resulting equation as a function of the unknown parameter $\sigma$:
\begin{eqnarray}
&&\setminus std\left(\left[\left(\frac{\Delta I}{I}\right)_2-2\frac{\Delta N}{N}\right]\right)-\nonumber\\
&&+\setminus abs\left(\frac{G(\sigma,(\Delta\omega)_2)}{[G(\sigma,(\Delta\omega)_2)-G(\sigma,(\Delta\omega)_1)]}\right)\times\nonumber\\
&&\times\setminus std\left(\left[\left(\frac{\Delta I}{I}\right)_2-\left(\frac{\Delta I}{I}\right)_1\right]\right)=0\label{equ11}
\end{eqnarray}
where the symbols $\setminus std$ and $\setminus abs$ in the previous equation are indicating the operations of standard deviation and absolute value, respectively.
\begin{figure}[!tbh]
    \centering
    \includegraphics[width=0.4\textwidth]{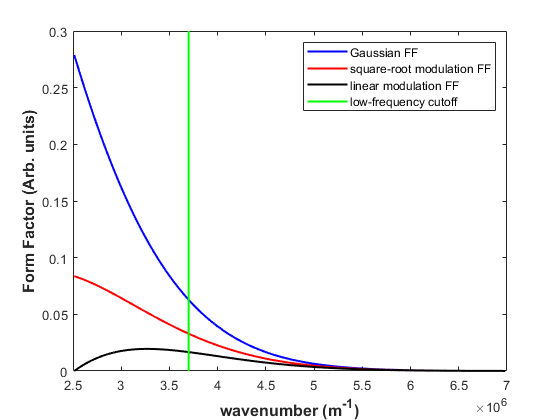}
    \caption{Pure, square-root and the linear modulated Gaussian FF [$\sigma=1.5$ fs (rms)] in the wavelength band 0.9-2.5 $\mu m$ of the ECOL spectrometer. The vertical line marks the wavelength band 0.9-1.7 $\mu m$, see also fig.(\ref{Orlandi:fig5}).}
    \label{Orlandi:fig3}
\end{figure}
It is worth noting that the arguments of the operator standard deviation $\setminus std$ in eq.(\ref{equ11}) are adimensional and consequently addable. The capability of eq.(\ref{equ11}) to predict the absolute value of the bunch length $\sigma$ was numerically tested in relation to the ECOL-BCM of SwissFEL for a Gaussian FF with $\sigma$ ranging from 0.5 to 3.0 fs, step 0.5 fs. We supposed a uniform transfer function for the ECOL-BCM pyrodetector in the wavelength band 0.9-4.0 $\mu m$, while, for the spectrometer, we supposed a transfer function either uniform or affected by a low-frequency modulation cutoff $M(\omega)$ in the wavelength bands 0.9-2.5 $\mu m$ and 0.9-1.7 $\mu m$, respectively, see fig.(\ref{Orlandi:fig3}). In particular, we considered a square-root ($\alpha=0.5$, $M_{min}=0.3$) and a linear ($\alpha=1.0$, $M_{min}=0.0$) modulation function of the spectrometer response, see note \cite{formula}. In fig(\ref{Orlandi:fig3}), the pure, square-root and the linear modulated Gaussian FF are plotted for a $\sigma=1.5$ fs. For each $\sigma$, relative statistical fluctuations of the two ECOL-BCM signals $\left(\frac{\Delta I}{I}\right)_i$ - with $i=1,2$ - were calculated from randomly extracted sequences of $\frac{\Delta\sigma}{\sigma}$ with rms deviation 1$\%$. Similarly, sequences of $\frac{\Delta N}{N}$ with rms deviation of 1$\%$ were randomly extracted.
\begin{figure}[!tbh]
    \centering
    \includegraphics[width=0.4\textwidth]{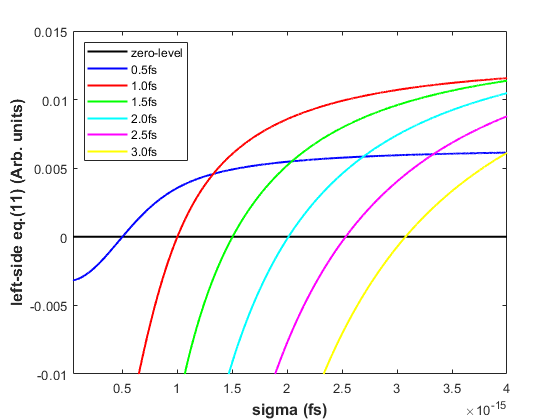}
    \caption{Test results of eq.(\ref{equ11}) from ECOL-BCM: the abscissae of the curves intercepts with the "zeros-level" line are the model-estimated absolute values of $\sigma$. Simulation settings: pyrodetector and spectrometer with wavelength bands 0.9-4.0 $\mu m$ and 0.9-2.5 $\mu m$, respectively; Gaussian FF with $\sigma$=0.5-3.0 fs; model-input $\frac{\Delta\sigma}{\sigma}$ and $\frac{\Delta N}{N}$ with rms deviation of 1 $\%$.}
    \label{Orlandi:fig4}
\end{figure}
The so obtained sequences of $\left(\frac{\Delta I}{I}\right)_i$ - with $i=1,2$ - and $\frac{\Delta N}{N}$ were used as input variables of eq.(\ref{equ11}). The absolute values of the bunch length were finally obtained by calculating the "zeros" of eq.(\ref{equ11}) as a function of the test variable $\sigma$ as shown in fig.(\ref{Orlandi:fig4}) where the case of a pure Gaussian FF over the full wavelength band (0.9-2.5 $\mu m$) of the ECOL spectrometer was considered. The abscissae of the curve intercepts with the "zeros" level in fig.(\ref{Orlandi:fig4}) permit to determine - eq.(\ref{equ11}) -  the electron bunch length within an error from 0.5$\%$ to 2.5$\%$, see "blue" curve ("Gaussian FF") in fig.(\ref{Orlandi:fig5})(a) compared to the "black" curve ("reference").
\begin{figure}[!tbh]
    \centering
    \includegraphics[width=0.4\textwidth]{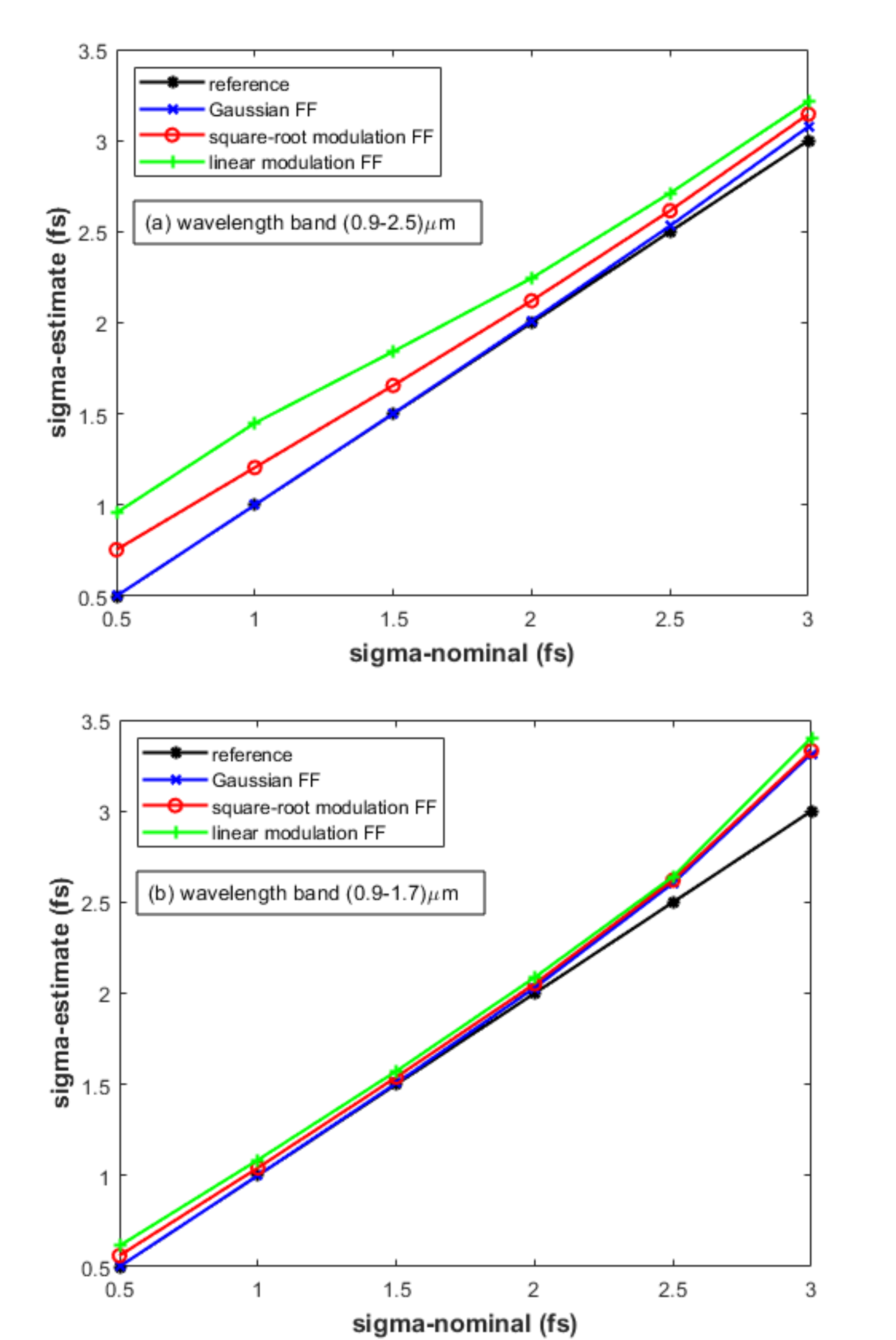}
    \caption{Test results of eq.(\ref{equ11}) from ECOL-BCM: model-estimated vs model-input values of $\sigma$. Simulation settings: Gaussian FF with $\sigma$=0.5-3.0 fs filtered by a uniform, square-root and linear modulated spectrometer response in the wavelength bands (a) 0.9-2.5 $\mu m$ and (b) 0.9-1.7 $\mu m$; model-input sequences of $\frac{\Delta\sigma}{\sigma}$ and $\frac{\Delta N}{N}$ with rms deviation of 1 $\%$. The "blue" curve in fig.5 (a) - Gaussian FF -  draws the abscissae of the curve intercepts to the "zero-level" as shown in fig.(\ref{Orlandi:fig4}).}
    \label{Orlandi:fig5}
\end{figure}
For the aforementioned case of the ECOL-BCM, under the hypothesis of a pure, square-root and linear frequency-modulated Gaussian FF - see fig.(\ref{Orlandi:fig3}) - in the spectrometer wavelength bands 0.9-2.5 $\mu m$ and 0.9-1.7 $\mu m$, the complete overview of the absolute determination of the bunch length $\sigma$ based on the "zeros" calculation of eq.(\ref{equ11}) is shown in Fig.(\ref{Orlandi:fig5}) for model-input values of $\sigma$ in the range 0.5-3.0 fs. With the exclusion of the less significative values 0.5 and 3.0 fs - the ECOL-BCM design is indeed optimized for the bunch length interval 0.7-3.0 fs \cite{Frei} - the error on the model-predicted $\sigma$ is in the range 20-5$\%$ and 5-3$\%$ in the case of a square-root frequency modulation of the Gaussian FF in the spectrometer wavelength bands 0.9-2.5 $\mu m$ and 0.9-1.7 $\mu m$, respectively, see fig.(\ref{Orlandi:fig5}). Whereas, in the case of a linear modulation of the Gaussian FF in the spectrometer wavelength band 0.9-1.7 $\mu m$, the error on the model-predicted $\sigma$ stays in the range 8-4$\%$, see fig.(\ref{Orlandi:fig5})(b). In conclusion, as already mentioned, thanks to the processing of the relative variations of the input signals, the model-prediction - eq.(\ref{equ11}) - of the absolute value of the bunch length $\sigma$ shows robustness and weakly dependency on unexpected or unknown bunch-length independent frequency modulation of the FF.

We presented a mathematical model that permits to track the relative variation of the bunch length $\frac{\Delta\sigma}{\sigma}$ from the processing of the shot-sequential statistical fluctuations of a BCM and a charge monitor signals. Moreover, in the case of a BCM detecting the radiation pulse simultaneously in two distinct wavelength bands, the proposed method also permits an absolute determination of the electron bunch length. The presented results pave the way to a fine tuning of the machine compression via a BCM based feedback loop driven by the absolute bunch-length measurement instead of a bunch-length dependent signal. Finally, for a CW-linac driven FEL, the proposed method discloses evident perspectives of a fully non-invasive and absolute determination of the electron bunch length.


\begin{thebibliography}{99}


\bibitem{LCLS} P. Emma, et al., "First lasing and operation of an ångström-wavelength free-electron laser", Nat. Photon. 4, 641–647 (2010).

\bibitem{SACLA} T. Ishikawa, et al., "A compact X-ray free-electron laser emitting in the sub-ångström region", Nat. Photon. 6, 540–544 (2012).

\bibitem{PAL} H.-S. Kang, et al., "Hard X-ray free-electron laser with femtosecond-scale timing jitter", Nat. Photon. 11, 708–713 (2017).

\bibitem{EuropXFEL} M. Altarelli, "The European X-ray free-electron laser facility in Hamburg", Nucl. Instrum. Methods B 269, 2845–2849 (2011).

\bibitem{atomic1} C. Bostedt, "Linac Coherent Light Source: the first five years", Rev. Mod. Phys. 88, 015007 (2016).

\bibitem{Milne}
C. J. Milne et al., SwissFEL: The Swiss X-ray Free Electron Laser, Appl. Sci. 7(7), 720 (2017).
\url{https://doi.org/10.3390/app7070720}

\bibitem{atomic2} P. Skopintsev et al., "Femtosecond-to-Millisecond Structural Changes in a Light-Driven Sodium Pump", Nature 583, 314 (2020).

\bibitem{atomic3} C. Bacellar et al., "Spin Cascade and Doming in Ferric Hemes: Femtosecond X-Ray Absorption and X-Ray Emission Studies", Proc. Natl. Acad. Sci. U.S.A. 117, 21914 (2020).

\bibitem{chemich1} M. Dell’Angela, et al., "Real-time observation of surface bond breaking with an X-ray laser", Science 339, 1302–1305 (2013).

\bibitem{chemich2} H. Oeström, et al., "Probing the transition state region in catalytic CO oxidation on Ru", Science 347, 978–982 (2015).

\bibitem{biological1} C. Bacellar et al., "Femtosecond X-Ray Spectroscopy of Haem Proteins", Faraday Discuss. 228, 312 (2021).

\bibitem{biological2} P. B{\"a}th et al., "Lipidic Cubic Phase Serial Femtosecond Crystallography Structure of a Photosynthetic Reaction Centre", Acta Crystallogr D Struct Biol 78, 698 (2022).

    \bibitem{Wu}
J. Wu , P. Emma,
"Linac Coherent Light Source (LCLS) Bunch-Length Monitor using Coherent Radiation"
Contributed to 2006 Linear Accelerator Conference, 21-25 August 2006, knoxville, TN, USA, (2006) SLAC-PUB-12121.

    \bibitem{Loos}
H. Loos , T. Borden, P. Emma, J. Frisch, J. Wu,
"Relative Bunch Length Monitor for the Linac Coherent Light Source (LCLS) using Coherent Edge Radaiation",
Proceedings of PAC07, Albuquerque, New Mexico, USA, FRPMS071, 4189-4191 (2007).
\url{https://accelconf.web.cern.ch/p07/PAPERS/FRPMS071.PDF}


\bibitem{Frei2}
F. Frei, I. Gorgisyan, B. Smit, G.L. Orlandi, B. Beutner, E. Prat, R. Ischebeck,
V. Schlott, "Development of Electron Bunch Compressor Monitors for SwissFEL",
Proceedings of IBIC2013, Oxford, UK, 769-771, WEPC36 (2013).
\url{accelconf.web.cern.ch/IBIC2013/papers/wepc36.pdf}

\bibitem{Frei}
	F. Frei, R. Ischebeck,
	"Electron bunch compression monitors for short bunches - commissioning results from SwissFEL",
	in \emph{Proc. IBIC2019},
	Malmoe, Sweden, Sep. 2019, pp. 578--581.
	\url{doi:10.18429/JACoW-IBIC2019-WEPP026}

\bibitem{Lockmann}
    N. M. Lockmann, C. Gerth, B. Schmidt, and S. Wesch, “Noninvasive THz spectroscopy for bunch current profile reconstructions at MHz repetition rates”, Phys. Rev. Accel. Beams, 23, 112801 (2020). doi: 10.1103/PhysRevAccelBeams.23.112801.
    \url{https://doi.org/10.1103/PhysRevAccelBeams.23.112801}

\bibitem{Gerth}
    Ch. Gerth, N. M. Lockmann,
    "Bunch Compression Monitor based on Coherent Diffraction Radiation at European XFEL and FLASH",
    Proc. IBIC2021, Pohang, Rep. of Korea, 400-403 (2021).
    \url{doi:10.18429/JACoW-IBIC2021-WEPP14}

\bibitem{Williams}
G.P. Williams, C.J. Hirschmugl, E.M. Kneedler, P.Z. Takacs, M. Shleifer, Y.J. Chabal, F.M. Hoffmann,
"Coherence Effects in Long-Wavelength Infrared Synchrotron Radiation Emission",
Phys. Rev. Lett. 62, 261 (1989).
\url{https://doi.org/10.1103/PhysRevLett.62.261}

\bibitem{Ishi}
K. Ishi, Y. Shibata, T. Takahashi, H. Mishiro, T. Ohsaka, M. Ikezawa, Y. Kondo, T. Nakazato, S. Urasawa, N. Niimura, R. Kato, Y. Shibasaki, M. Oyamada,
"Spectrum of coherent synchrotron radiation in the far-infrared region"
Phys. Rev. A 43, 5597 (1991).
\url{https://doi.org/10.1103/PhysRevA.43.5597}

\bibitem{Happek}
U. Happek, A.J. Sievers, E.B. Blum,
"Observation of coherent transition radiation"
Phys. Rev. Lett. 67, 2962 (1991).
\url{https://doi.org/10.1103/PhysRevLett.67.2962}

\bibitem{Shibata2}
Y. Shibata, K. Ishi, T. Takahashi, T. Kanai, M. Ikezawa, K. Takami, T. Matsuyama, K. Kobayashi, and Y. Fujita,
"Observation of coherent transition radiation at millimeter and submillimeter wavelengths",
Phys. Rev. A 45, R8340(R) (1992).
\url{https://doi.org/10.1103/PhysRevA.45.R8340}

\bibitem{Shibata}
Y. Shibata, S. Hasebe, K. Ishi, T. Takahashi, T. Ohsaka, M. Ikezawa, T. Nakazato, M. Oyamada, S. Urasawa, T. Yamakawa, Y. Kondo,
"Observation of coherent diffraction radiation from bunched electrons passing through a circular aperture in the millimeter- and submillimeter-wavelength regions"
Phys. Rev. E 52, 6787 (1995).
\url{https://doi.org/10.1103/PhysRevE.52.6787}

\bibitem{Cast}
    M. Castellano, V. A. Verzilov, L. Catani, A. Cianchi, G. Orlandi, M. Geitz
    "Measurements of coherent diffraction radiation and its application for bunch length diagnostics in particle accelerators",
    Phys. Rev. E 63, 056501 (2001)
    \url{https://doi.org/10.1103/PhysRevE.63.056501}

\bibitem{Veronese}
    M. Veronese, R. Appio, P. Craievich, G. Penco
    "Absolute Bunch Length Measurement Using Coherent Diffraction Radiation"
    Phys. Rev. Lett. 110, 074802 (2013).
 \url{https://doi.org/10.1103/PhysRevLett.110.074802}

\bibitem{SFCDR}
SwissFEL Conceptual Design Report, PSI Bericht Nr. 10-04
April 2012.

\bibitem{TS}
T. Schietinger et al.,
"Commissioning experience and beam physics measurements at the SwissFEL Injector Test Facility",
Phys. Rev. Accel. Beams 19, 100702 (2016).
\url{https://doi.org/10.1103/PhysRevAccelBeams.19.100702}

\bibitem{Paraliev}
M. Paraliev et al.,
"SwissFEL double bunch operation",
Phys. Rev. Accel. and Beams 25, 120701 (2022)
\url{https://doi.org/10.1103/PhysRevAccelBeams.25.120701}

\bibitem{Prat}
E. Prat et al.,
“A compact and cost-effective hard X-ray freeelectron laser driven by a high-brightness and low-energy electron beam”,
Nature Photonics, vol. 14, pp. 748–754, 2020.
\url{https://doi.org/10.1038/s41566-020-00712-8}

\bibitem{Abela}
R. Abela, et al.,
"The SwissFEL soft X-ray free-electron laser beamline: Athos",
J. Synchrotron Radiat. 26, 1073-1084 (2019).
\url{https://doi.org/10.1107/S1600577519003928}

\bibitem{Nodvick}
J. S. Nodvick, D. S. Saxon,
"Suppression of Coherent Radiation by Electrons in a Synchrotron",
Phys. Rev. 96, 180 (1954).
\url{https://doi.org/10.1103/PhysRev.96.180}

\bibitem{Hirschmugl}
C. J. Hirschmugl, M. Sagurton, G. P. Williams
"Multiparticle coherence calculations for synchrotron-radiation emission",
Phys. Rev. A 44, 1316 (1991).
\url{https://doi.org/10.1103/PhysRevA.44.1316}

\bibitem{Bosch}
    R.A. Bosch
    "Edge radiation in an electron storage ring"
    Il Nuovo Cimento D volume 20, 483–493 (1998)
    \url{https://doi.org/10.1007/BF03185543}

\bibitem{Bosch2}
R. A. Bosch,
"Extraction of edge radiation within a straight section of Aladdin",
Rev Sci Instrum 73, 1423–1426 (2002).
\url{https://doi.org/10.1063/1.1435813}

\bibitem{Orlandi}
    G.L. Orlandi
    “Misura della lunghezza del pacchetto di elettroni dell'acceleratore
    lineare superconduttivo TTF (tesla test facility) con la radiazione di transizione e di
    diffrazione”,
    Tesi di dottorato XII ciclo (2000), TDR 2000 000337, Biblioteca Nazionale Centrale di Firenze.
    \url{https://opac.bncf.firenze.sbn.it/bncf-prod/resource?uri=TSI0000337&v=l&dcnr=0}

\bibitem{GR}
    I.S. Gradshteyn and I.M. Ryzhik,
    “Table of Integrals, Series, and Products”, Fifth Edition, A. Jeffrey Editor, Academic Press (1994).

    \bibitem{Emma}
    P. Emma, J. Frisch, P. Krejcik,
    "A Transverse RF Deflecting Structure for Bunch Length and Phase Space Diagnostics", (2000) LCLS-TN-00-12.

\bibitem{TDC1} R. Akre, L. Bentson, P. Emma, and P. Krejcik, Proceedings of the Particle Accelerator Conference (PAC 2001), Chicago, IL, 2001 (IEEE, New York, 2001).

\bibitem{TDC2} D. Alesini, G. Di Pirro, L. Ficcadenti, A. Mostacci, L. Palumbo, J. Rosenzweig, and C. Vaccarezza, Nucl. Instrum. Methods Phys. Res., Sect. A 568, 488 (2006).

\bibitem{TDC3} M. Röhrs, C. Gerth, H. Schlarb, B. Schmidt, and P. Schmüser, Phys. Rev. ST Accel. Beams 12, 050704 (2009).

\bibitem{TDSInj} P. Craievich, R. Ischebeck, F. L\"ohl, G.L. Orlandi, E. Prat, \textit{Transverse Deflecting Structures for Bunch Length and Slice Emittance Measurements on SwissFEL}, Proceedings of FEL2013, New York, NY, USA, TUPSO14 236-241 (2013).

\bibitem{formula}
Mathematical note, hypothetic low-frequency modulation function of the ECOL-BCM spectrometer response:
\begin{eqnarray}
M(\omega)=\left(\frac{1-M_{min}}{\omega_{max}^\alpha-\omega_{min}^\alpha}\right)\omega^\alpha+\frac{\omega_{max}^\alpha M_{min}-\omega_{min}^\alpha}{\omega_{max}^\alpha-\omega_{min}^\alpha}\nonumber
\end{eqnarray}

\end{thebibliography}
\end{document}